\documentstyle[twocolumn]{article}

\title{\bf On the Design of an Expert Help System  \\ 
for Computer Algebra Systems\thanks{This is a short, less technical version of [1].}\ \thanks{This research was supported by Conselho 
Nacional de Desenvolvimento Cient\'\i fico e Tecnol\'ogico - CNPq.}}
\author{R. P. dos Santos\thanks{On leave of Centro Brasileiro de
Pesquisas F\'\i sicas, Rua Xavier Sigaud, 150, 22290 Rio de Janeiro, RJ, Brazil.} \ and W. L. Roque\thanks{On leave of Universidade de Bras\'\i lia, Instituto de Ci\^encias Exatas, 70910 Bras\'\i lia, DF, Brazil.} \\ 
                 Universit\"at Karlsruhe \\ 
        Institut f\"ur Algorithmen und Kognitive Systeme \\ 
         Am Fasanengarten 5 -- D7500 Karlsruhe 1 -- FRG \\
         Phone: (+49) 721 6084328 \\ 
         BitNet: kg07@dkauni2 and kg03@dkauni2}
\date{}
     
\begin{document}

\setlength{\oddsidemargin}{0cm}
\setlength{\evensidemargin}{0cm}
\setlength{\textwidth}{17cm}
\setlength{\topmargin}{-1cm}
\setlength{\headheight}{0cm}
\setlength{\headsep}{0cm}
\setlength{\textheight}{22,5cm}

\maketitle

\section{Introduction}

REDUCE\cite{hearn88,maccallum90}, like many modern Computer Algebra Systems (CAS) (MACSYMA, MATHEMATICA, MAPLE, SCRATCHPAD to name a few), embodies a large amount of mathematical knowledge which is spread out through thousands of procedures in the source code of the system. Most of this knowledge is, however, in an {\em implicit} form almost inaccessible to the user, who would have to decipher the source files to recover it. Also, to profit from it one needs to know  in addition the capabilities of these computing systems and how to operate them.

The process of learning how to use a CAS may be done by reading throughout the manual available for the system, a sort of user's guide, a book or in the practical {\em hands on} approach. However, soon the begginers get stuck and the most confortable and easiest way to solve the problem is consulting an {\em expert} on the system. They, in general, do not want to waste their time reading either a book or manual looking for the terminological structure of the system.  Particularly, to its semantics, syntax and/or examples just to be able to start using the system to solve a simple problem: an integral, for instance. That is, perhaps, the 
reason why many the potential users with no previous experience with computers keep avoiding to use them.

These kinds of informations could be easily provided by an
on-line help system (and some progress in this direction have been attained\cite{rand90}) or even better, by an Intelligent Tutorial System\cite{sleeman82}. The idea behind is that existing a stored knowledge base, consisting of a large number of information elements, structured in levels of specialization, an {\em expert help system} could generate by inference (not just by pure recovering) an information, which might not be explicity stored and that might be an adequate (in terms of abstraction, difficulty, detail, etc.) directive to the user. An intelligent help facility could reason about the
user's query and then give a reasonable reply or else, suggest what has to be consulted or even what to do next.

It is our intention here only to discuss the nature, complexity and tools concerning the design of {\em Smart Help}, an expert help system with these features. Presently, the {\em Smart Help} domain knowledge base concerns REDUCE, however it may well be used to implement different CAS bases. Since the hybrid knowledge
representation system (KRS) MANTRA\cite{bittencourt90a} can be seen as a knowledge representation shell, one could make use of its (formal) reasoning facilities to build the {\em Smart Help} system. The ideas forwarded herein are in a prototypal basis, but we hope that they will raise the interest of the CAS community and evolve to reach at the end the full system implementation.

In section \ref{taxonomy} we comment briefly on the computer algebra system REDUCE and propose a taxonomy for the knowledge
embodied by this system. In section \ref{mantra} we describe the knowledge representation system MANTRA. Section \ref{shelp} and 5 are concerned with the {\it Smart Help} system design.  Finally, in section 6, we give some conclusions and suggest further upgrades which would turn {\em Smart Help} into a truly {\em Intelligent Tutorial System}\cite{sleeman82}. An example of the interaction is also given. Further details can be found in \cite{dossantos90}.

\section{The CAS REDUCE}

REDUCE\cite{hearn88} is a fairly powerful system for carrying out a variety of mathematical calculations such as to manipulate polynomials in a variety of forms, simplify expressions, to differentiate and integrate algebraic expressions, do some modern differential geometry calculations, study the Lie-symmetries of systems of partial differential equations, and many others.

This system is made out of different knowledge domains (mathematical, terminological, computational, etc.). In order to represent in {\em Smart Help} the specific knowledge domain involved in operating REDUCE we propose to classify it in the following categories:

\label{taxonomy}
\begin{description}
\item[syntax:] Informations about the number and type of the arguments, and requisites and prerequisites of a REDUCE command, declaration or operator (that is the only type of information given by many help systems).
\item[terminology:] Definitions of the various terms like {\em identifier}, {\em operator}, {\em kernel}, etc., employed in documents related to REDUCE and, also important, in error messages from REDUCE (this matter can be quite confusing to the initial user).
\item[concepts:] Informations like the fact that {\em integration}
in REDUCE is represented by an operator called {\bf INT}, or a reference for the algorithm which it implements.

\item[procedures:] General sequences of commands which should be
given to perform a certain task like defining a new infix operator: one has to declare the operator infix through the INFIX declaration and then give him a precedence by means of the PRECEDENCE declaration.
\item[heuristics:] General rules and tips that simplify and improve approaches to problem-solving (this kind of information comes with
experience and is one of the most frequent in consultations from beginners).  For example, ``If you want to compute a definite integration, you can try evaluating the indefinite integral, saving the resulting expression in a variable and then substituting locally the limits of integration in it and finally subtracting the results''.
\end{description}

\section{The KRS MANTRA\protect\footnotemark} \footnotetext{This description of the MANTRA System is based on \cite{bittencourt90a}} \label{mantra} 

MANTRA\cite{bittencourt90a} is a hybrid
 knowledge representation system which integrates three different knowledge representation methods: 
\begin{description}
\item[Four-valued first-order logic language,] \qquad used to express {\em assertional knowledge}, which is decidable but presents a weaker entailment mechanism which excludes the chaining of independent facts, thus ruling out {\em modus ponens} but allowing quantifiers.
\item[Terminological language] (a kind of {\em frame} method), extended to allow the definition of concepts and n-place relations over themselves.
\item[Semantic network,] which is a representation to define hierarchies with exceptions.
\end{description}

One could think that classical logic would be able to generate
all knowledge entailed (i.e., implicity represented) by a given concept. Unfortunately, however, the systems based on the complete first-order logic, suffer from the inherent problem of {\em Combinatorial Explosion}. Also, the entailment problem is not decidable in first-order logic. In addition, the knowledge to be represented is often incomplete and incoherent.

Due to these facts, the knowledge representation system MANTRA has been designed associating the three knowledge representation
methods above, based on a common four-valued semantic, which allows the modeling of the aspects of {\em ignorance} and {\em inconsistency}, useful for representing incomplete and/or incoherent knowledge.

\section{The Conception of the {\em Smart Help} Expert System}\label{shelp}

The conception of {\em Smart Help} follows the tradition of help systems being passive. This means that the user learns how to use REDUCE playing freely with it without being interrupted by the {\em Smart Help}. To the user it looks like a normal REDUCE session but MANTRA is running behind and is accessible as an operator, by means of the interface MANTRA-REDUCE. When the user gets a confusing answer or a meaningless error message (and in fact REDUCE users know how often this happens), or even when he does not know what to do next to get his calculations done, he invokes the {\em Smart Help} to clarify the point.

The problem might have been caused by earlier mistakes (a forgotten variable definition long before, for example). An explanation facility\cite{marti84} to trace the session history and find the exact place at which the misconception first had its effect was not included in {\em Smart Help}. It is then left to the user to do the right question to get the right answer. That is, progress can be made only if, from the correct definition, usage, prerequisites, etc., of the queried topic, as returned by the {\em Smart Help}, the user can pinpoint the misconception which caused the problem.

\section{The Architecture of {\em Smart Help}}

Technically, {\em Smart Help} is a Production System on the top of a particular implementation of MANTRA which has REDUCE integrated as an additional knowledge representation module. Since the heuristic level of MANTRA has not yet been implemented, being presently represented by the Lisp language itself, {\em Smart Help} is coded in Lisp and resides in the same Lisp session of MANTRA. 

Considering the taxonomy of the knowledge embodied by REDUCE, presented in section \ref{taxonomy} above, and the knowledge representation methods available in MANTRA, as described previously in
section \ref{mantra}, we had to find the best fit of both to guarantee efficiency in recovering knowledge from the base and in reasoning with it, according to the inherent structure of each knowledge category and to its adaptiveness to the specific representation method. The five categories of knowledge were implemented as {\em aspects} of the knowledge base {\em object},  making use of Corbit\cite{desmedt87}, an object-oriented extension of Common Lisp. Details of this can be found in \cite{dossantos90}.

A number of rules were implemented in the production system of {\em Smart Help}. We present them in the following. Presently, they are inserted in the code itself. We consider now to get them defined in a production rule base, what would give flexibility and clearness to our system.

When asked by the user about a topic, {\em Smart Help} 
\begin{enumerate}
\item Assumes that the present level of familiarity of the user with REDUCE (student model) can be inferred from the level of specification of the queried topic, which is characterized by a numeric heuristical parameter, ranging from 1 to 3, associated to it. For example if someone queries about ``integration'' (very general -- parameter = 1) it seems probable to be a very novice user but if one queries about ``infix-operators'' (more specialized -- parameter = 2) it should be considered as an user with some familiarity. 
\item Queries the domain knowledge base, to recover all informations
related to the given topic and store them in the ``answer'' structure. This process consists in querying the {\em conceptual}, {\em terminological}, {\em syntactical}, {\em procedural} and {\em
heuristical} aspects of the base.
\item Evaluates the level of generality of the terms recovered to see if they are in the same level of specialization (same value of the parameter). If a concept is too specific (much greater value of the parameter), {\em Smart Help} tries to redefine it in terms of more
general concepts.  If a concept is too general, however, it is simply  deleted.
\item Formats the recovered knowledge in the form of a readable answer, defining the queried concept and its usage in terms of the recovered knowledge. Presently this process is quite crude as it is a peripherical point of the implementation. It will be improved latter on.
\item Prints the answer returning control to REDUCE.
\end{enumerate}

For better understanding, suppose that an user with no experience with REDUCE wants to do some calculations, for instance, integrate an expression in terms of a certain variable. Having access to an initialized session of REDUCE (in which the heading shows how to invoke {\em Smart Help}), the interaction would consist in typing {\tt shelp integration} , and {\em Smart Help} would promptly reply: \\
\begin{verse}
{\tt "SHelp - Version 2.0: 23 Aug 1990"} \\ 
{\tt INDEFINITE-INTEGRATION is the default for INTEGRATION.} \\ 
{\tt INTEGRATION is represented in Reduce by INT.} \\ 
{\tt Its syntax is:} \\ 
{\tt INT(scalar-expression,variable)} \\ 
{\tt Ex.: INT(LOG(X),X); } \\
{\tt INT is implemented through the } \\
{\tt SIMPINT-PROCEDURE-IN-INT-SOURCE-FILE.} \\
{\tt For DEFINITE INTEGRATION, one may try} \\
{\tt  to DO INDEFINITE-INTEGRATION,} \\
{\tt  to SAVE-RESULT-IN VARIABLE,} \\
{\tt  to LOCALLY-SUBSTITUTE VALUES.} \\ 
{\tt References: REDUCE MANUAL SECTION 7.4.} \\ 
{\tt See also: DERIVATION} 
\end{verse}

\section{Conclusions}

We have presented and proposed in this paper a fairly general design of an expert help facility for aiding users of Computer Algebra Systems. Although the expert help system presented here has been particularly oriented to REDUCE (as a consequence of our former experience with this system), we point out that the concept of {\em Smart Help} can be extended to other Computer Algebra Systems.

The reasons for introducing {\em Smart Help} facility include: 
\begin{itemize}
\item It will provide an on-line help for the system, aiding the users to find specific informations about the system terminology, structure, syntax, etc.
\item It will allow the potential user to access the whole capabilities of the system.  
\item It can contribute to the development of Intelligent Computer Algebra Systems\cite{calmet87b}, which more than simply being able to do calculations, could interact with the user and free him of many details concerning the specification of his problem.
\end{itemize}

The {\em Smart Help} has no intention to {\em teach} the user how to program efficiently in REDUCE or behave like a tutoring system.
However, it can be used in the learning process as a complimentary teaching tool.

Following the ideas addressed in this paper, we intend afterwards to develop an {\em Intelligent Tutorial System}\cite{sleeman82} (ITS) for REDUCE. The ITS should inherit all the compatible facilities already available in the {\em Smart Help}. In addition, many other facilities would become available, such as, a deeper understanding of REDUCE's semantics, keeping track of history, explanations\cite{marti84}, a dynamical reasonning on the student's model\cite{sleeman82}, etc.

A prototype of {\em Smart Help} is now running on a SUN work-station on an experimental basis. The full implementation of {\em Smart Help} as a final product was not our main concern here. This task will certainly need few more people working in a close colaboration to build up a satisfactory knowledge base to reach at the end the principal objective that is helping a CAS user. 

\section{Acknowledgement}

We would like to thank Prof. J. Calmet for enlightening discussions and for the warm hospitality provided by him and by all 
members of his group and the Conselho Nacional de Desenvolvimento Cient\'\i fico e Tecnol\'ogico -- CNPq for the financial support.

\end{document}